\documentclass[final,1p,times,twocolumn]{elsarticle}

\usepackage{amssymb}
\usepackage{amsfonts}
\usepackage{graphicx}
\usepackage{amsmath}

\newcounter{bla}

\journal{Computer Physics Communications}

\begin{document}

\begin{frontmatter}



\title{Linearized self-consistent quasiparticle GW method: Application to semiconductors and simple metals}

\author{A.~L.~Kutepov$^{a,}$\corref{a5}}
\author{V.~S.~Oudovenko$^{a}$\corref{a}}
\author{G.~Kotliar$^{a,b}$\corref{b}}

\cortext[a5] {Corresponding author.\\\textit{E-mail address:} kutepov@physics.rutgers.edu}
\address{$^{a}$Department of Physics and Astronomy, Rutgers University, Piscataway, NJ 08856}
\address{$^{b}$Brookhaven National Laboratory, NY}

\begin{abstract}
We present a code implementing the linearized self-consistent quasiparticle
GW method (scQPGW) in the LAPW basis. 
Our approach is based on the linearization of the self-energy around zero frequency which differs it from the existing implementations of the scQPGW method. 
The linearization allows us to use Matsubara’s frequencies instead of real ones. As a result it gives us an advantage in terms of efficiency, allowing us easily switch to the imaginary time representation the same way as in the space time method. The all electron LAPW basis set eliminates the need for pseudopotentials. We discuss the advantages of our approach, such as its $N^{3}$ scaling with the system size, as well as its shortcomings.

We apply our approach to study electronic properties of selected semiconductors, insulators, and simple metals and show that our code produces results very close to
the previously published scQPGW data. Our   implementation is a good platform for further many body  diagrammatic resummations such as GW+DMFT. 
\end{abstract}



\end{frontmatter}



{\bf PROGRAM SUMMARY/NEW VERSION PROGRAM SUMMARY}

\begin{small}
\noindent
{\em Program Title: LqsgwFlapw}                                          \\
{\em Journal Reference:}                                      \\
{\em Catalogue identifier:}                                   \\
{\em Licensing provisions: GNU General Public License}                                   \\
{\em Programming language:} Fortran 90                                   \\
{\em Computer:} Windows workstations, Linux/UNIX servers/workstations or clusters                                              \\
{\em Operating system: LINUX, MAC OS X, Windows}                                       \\
{\em RAM: 2-10} gigabytes per CPU (depending on system size).                                             \\
{\em Number of processors used: 1-192}                              
{\em Classification: 7.3}                                         \\
{\em External routines/libraries:}  BLAS, LAPACK, MPI(optional)                          \\
{\em Nature of problem:}\\
   Direct implementation of GW method scales as $N^4$ with the system size, which quickly becomes prohibitively time consuming even for the modern computers.\\
{\em Solution method:}\\
   We implemented GW method using the fact that some operations are better to perform in real space, whereas other are more computationally efficient in the reciprocal space. This makes our approach scale as $N^{3}$. \\
{\em Restrictions:}\\
   Limiting factor usually is memory available in a computer. Memory 10 GB/core allows us to study systems up to 15 atoms per unit cell.\\
{\em Running time:}\\
   From 10 minutes up to a few days. (Depending on the system size.)\\
\end{small}


\section{Introduction}
\label{intr}
The so called GW method was originally proposed by Hedin\cite{pr_139_A796} and was first applied to real materials by Hybertsen and Louie\cite{prb_34_5390} and by Godby et al\cite{prb_37_10159}. There are many successful implementations of this method in open source codes using
plane waves basis sets such as BerkeleyGW \cite{cpc_183_1269}, ABINIT \cite{zk_220_558}, and West \cite{jctc_11_2680}. There are also  codes implementating GW  in all electron basis sets   such as
exciting \cite{jcm_26_363202} and SPEX \cite{prb_81_125102}.

There are numerous computational developments in this area (see for example \cite{prb_93_115203,prb_93_125210,arx_1501_03141} and references therein). For our present study we found particularly useful publication by Rieger et al on the space-time method\cite{cpc_117_211} and work by Ku and Eguiluz on the application of Matsubara time in GW calculations\cite{prl_89_126401}.

Our main goal in the present work is implementation and testing the self-consistent quasi-particle GW method (scQPGW) which is a promising tool for studying electronic structure of moderately correlated materials, atoms and molecules\cite{tcc_347_99}. Whereas the current method usually overestimates the widths of spectral features in materials (such as band gaps, band widths, and exchange splitting) it is generally 
more accurate than the local density approximation (LDA). scQPGW method also has clear  advantages in comparison with  another popular approximation of GW - one shot GW method (implemented after self-consistent LDA calculation),  as not  being dependent on the starting point.

All previous implementations of scQPGW method\cite{prb_76_165106,prl_99_246403,prb_92_041115} are based on real frequencies. We have found however, that similar results can be obtained with an
approach based on imaginary frequency representation. We exploit the fact that we can easily transform  functions from imaginary frequency to imaginary time (and back) to  enhance the computational efficiency. We use all-electron approximation (Linear Augmented Plane Wave method, LAPW) as a basis of our approach, thus avoiding the need in pseudopotentials.

Our implementation of the GW method was outlined in our previous work\cite{prb_85_155129}, where we presented the general scheme of the approach with detailed description of scGW method in its fully
relativistic form and the application of the scheme to Am and Pu. Also, we provided there the total energy evaluation which was used earlier in Ref.\cite{prb_80_041103}. In this work we present the non-relativistic version of scGW method with special emphasis on the scaling of the most time consuming steps of the algorithm. Particularly, we stress on the overall scaling $N^{3}$ with the system size which is similar to the scaling of LDA. In the end we present a few numerical examples obtained using quasiparticle approximation scQPGW. Our implementation of scQPGW approach was used recently in Ref.\cite{arx_1504_07569} in the combination of scQPGW and one-shot DMFT.

\section{The basics of GW method}
\label{GW}

Below we outline the basic formulae of  the method introduced
earlier in Ref.\cite{prb_85_155129}.  The input for every
iteration is the Green function $G$ which is renewed until the self consistency reached. Then we perform a few steps, calculating the intermediate functions such as:

polarizability
\begin{align}  \label{GW_1}
P(12;\tau)=-G(12;\tau)G(21;\beta-\tau),
\end{align}
screened interaction
\begin{align}  \label{GW_2}
W(12;\nu)=V(12)+\int d(34) V(13)P(34;\nu)W(42;\nu),
\end{align}
self energy 
\begin{align}  \label{GW_3}
\Sigma(12;\tau)=-G(12;\tau)W(21;\tau),
\end{align}
new Green's function
\begin{align}  \label{GW_4}
G(12;\omega)&=G_{0}(12;\omega)\nonumber\\&+\int d(34) G_{0}(13;\omega)\Sigma(34;\omega)G(42;\omega).
\end{align}

In scQPGW approach the last step is replaced with a special construction of quasiparticle Green's function, which was introduced in
\cite{prb_85_155129}

\begin{align}  \label{GW_5}
G\leftarrow G[G_{0};\Sigma].
\end{align}

The details of this step in our implementation of QPGW approach are given in section \ref{QP}.

\section{Representation of band states in LAPW method}
\label{LAPW}

In the LAPW method \cite{prb_12_3060} one represents the band states in the interstitial region as a linear combination of plane waves

\begin{align}  \label{LAPW_1}
\Psi^{\alpha\mathbf{k}}_{\lambda}(\mathbf{r})=\frac{1}{\Omega_{0}}\sum_{\mathbf{G}}A^{\alpha\mathbf{k}\lambda}_{\mathbf{G}}
e^{i(\mathbf{k}+\mathbf{G})\mathbf{r}},
\end{align}
where $\alpha$ is the spin index, $\mathbf{k}$ is the point in the Brillouin zone, $\mathbf{G}$ labels plane waves, and $\Omega_{0}$ being the volume of the unit cell. Index $\lambda$ stands for the band states, which in this work are the eigen states of an effective Hartree-Fock Hamiltonian constructed with the quasi-particle Greens function \cite{prb_85_155129}. Inside the MT-sphere at atom $\mathbf{t}$ it is convenient to represent band states as linear
combinations of orbitals belonging to that MT-sphere

\begin{align}  \label{LAPW_2}
\Psi^{\alpha\mathbf{k}}_{\lambda}(\mathbf{r})|_{\mathbf{t}}=\sum_{L}Z^{\alpha\mathbf{k}\lambda}_{\mathbf{t}L}
\varphi^{\alpha\mathbf{t}}_{L}(\mathbf{r}),
\end{align}
where index $L$ combines angular momentum indexes $l,m$ and any additional indexes to distinguish the orbitals inside the sphere (for example, it distinguishes between the solutions of radial equations and their energy derivatives). Representations
(\ref{LAPW_1}) and (\ref{LAPW_2}) will be used throughout the paper.

\section{Product basis conventions}
\label{PB_def}

We define product basis functions $M^{\mathbf{q}}_{K}(\mathbf{r})$ as plane waves in the interstitial region and optimized basis functions inside MT spheres.  We use the index $K$ to label the product basis functions  
 in all MT-spheres and in the interstitial region. When index $K$ runs over the functions inside MT-sphere $\mathbf{t}$ then

\begin{align}  \label{PB_1}
M^{\mathbf{q}}_{K}(\mathbf{r}+\mathbf{R})=\Big\{\begin{array}{cc}
                                            0 & \mathbf{r}\notin \mathbf{t} \\
                                            e^{i\mathbf{q}\mathbf{R}}M^{\mathbf{t}}_{K}(\mathbf{r}) & \mathbf{r}\in \mathbf{t}
                                          \end{array}
.
\end{align}

When it runs over the functions in the interstitial region, we associate the index with plane waves $\mathbf{G}_{K}$:

\begin{align}  \label{PB_1a}
M^{\mathbf{q}}_{K}(\mathbf{r}+\mathbf{R})=\Big\{\begin{array}{cc}
                                            e^{i\mathbf{q}\mathbf{R}}e^{i(\mathbf{q}+\mathbf{G}_{K})\mathbf{r}} & \mathbf{r}\in Int \\
                                            0 & \mathbf{r}\notin Int
                                          \end{array}
.
\end{align}

The word "optimized" means that we build the space of all products of orbitals $\varphi^{\alpha\mathbf{t}}_{L}$ in each MT-sphere and construct
linear independent ortho-normal basis in this space, following the Ref.\cite{prb_81_125102}.

The defined above product basis is not ortho-normal in the interstitial region. So we also define the dual product basis

\begin{align}  \label{PB_2}
\widetilde{M}^{\mathbf{q}}_{K}(\mathbf{r})=\Big\{\begin{array}{cc}
                                            \sum_{K'\in Int}\langle M^{\mathbf{q}}_{K'}|M^{\mathbf{q}}_{K}\rangle^{-1}M^{\mathbf{q}}_{K'} & \mathbf{r}\in Int \\
                                            M^{\mathbf{q}}_{\mathbf{t}K}(\mathbf{r}) & \mathbf{r}\in \mathbf{t}
                                          \end{array}
,
\end{align}
which is ortho-normal to the basis (\ref{PB_1}): $\langle \widetilde{M}^{\mathbf{q}}_{K'}|M^{\mathbf{q}}_{K}\rangle=\delta_{KK'}$.

We expand the polarizability in dual basis

\begin{align}  \label{PB_3}
P(\mathbf{r};\mathbf{r}';\tau)=\frac{1}{N_{\mathbf{k}}}\sum_{\mathbf{q}}\sum_{KK'}\widetilde{M}^{\mathbf{q}}_{K}(\mathbf{r})P^{\mathbf{q}}_{KK'}(\tau)
\widetilde{M}^{^{*}\mathbf{q}}_{K'}(\mathbf{r}'),
\end{align}
with $N_{\mathbf{k}}$ being the number of points in the Brillouin zone.

Correspondingly to express in product basis the interaction we have to calculate the integral

\begin{align}  \label{PB_4}
W^{\mathbf{q}}_{KK'}(\tau)=\int\int d\mathbf{r}d\mathbf{r}' \widetilde{M}^{^{*}\mathbf{q}}_{K}(\mathbf{r})W(\mathbf{r};\mathbf{r}';\tau)
\widetilde{M}^{\mathbf{q}}_{K'}(\mathbf{r}').
\end{align}

Because of the orthogonality of the original and dual product basis sets it is convenient to think of the interaction as an expansion in original
product basis set:

\begin{align}  \label{PB_5}
W(\mathbf{r};\mathbf{r}';\tau)=\frac{1}{N_{\mathbf{k}}}\sum_{\mathbf{q}}\sum_{KK'}M^{\mathbf{q}}_{K}(\mathbf{r})W^{\mathbf{q}}_{KK'}(\tau)
M^{^{*}\mathbf{q}}_{K'}(\mathbf{r}').
\end{align}

\section{Quasiparticle approximation}

\label{QP}

Different from the QPscGW method by Kotani et al.\cite%
{prb_76_165106}, our method is based exclusively on imaginary axis data.

We proceed as follows. In Dyson's equation for the Green function

\begin{align}  \label{qp_1}
G^{-1}_{\lambda\lambda^{\prime }}(\mathbf{k};\omega)=(i\omega+\mu-%
\varepsilon^{\mathbf{k}}_{\lambda})\delta_{\lambda\lambda^{\prime }} -\Sigma^{c}_{\lambda\lambda^{\prime }}(\mathbf{k};\omega),
\end{align}
where band indices $(\lambda ,\lambda ^{\prime })$ correspond to the effective exchange Hamiltonian\cite{prb_85_155129}, we approximate frequency
dependence of the self energy by a linear function

\begin{align}  \label{qp_2}
\Sigma^{c}_{\lambda\lambda^{\prime }}(\mathbf{k};\omega)=\Sigma^{c}_{\lambda%
\lambda^{\prime }}(\mathbf{k};\omega=0) +\frac{\partial\Sigma^{c}_{\lambda%
\lambda^{\prime }}(\mathbf{k};\omega)}{\partial (i\omega)}%
|_{\omega=0}(i\omega).
\end{align}
With this approximation the Dyson equation is simplified

\begin{align}  \label{qp_3}
G^{-1}_{\lambda\lambda^{\prime }}(\mathbf{k};\omega)=Z^{-1}_{\lambda%
\lambda^{\prime }}(\mathbf{k})(i\omega)+ (\mu-\varepsilon^{\mathbf{k}%
}_{\lambda})\delta_{\lambda\lambda^{\prime }} -\Sigma^{c}_{\lambda\lambda^{\prime }}(\mathbf{k};0),
\end{align}
where we have introduced a renormalization factor $Z$ 
matrix (not to be confused with the expansion coefficients in Eq. (\ref{LAPW_2})):

\begin{align}  \label{qp_4}
Z^{-1}_{\lambda\lambda^{\prime }}(\mathbf{k})=\delta_{\lambda\lambda^{\prime
}}-\frac{\partial\Sigma^{c}_{\lambda\lambda^{\prime }}(\mathbf{k};\omega)}{%
\partial (i\omega)}|_{\omega=0}.
\end{align}

Representing $Z$-factor as a symmetrical product

\begin{align}  \label{qp_5}
Z^{-1}_{\lambda\lambda^{\prime }}(\mathbf{k})=\sum_{\lambda^{\prime \prime
}}Z^{-1/2}_{\lambda\lambda^{\prime \prime }}(\mathbf{k})Z^{-1/2}_{\lambda^{%
\prime \prime }\lambda^{\prime }}(\mathbf{k}),
\end{align}
we reduce the Dyson equation to the following form

\begin{align}  \label{qp_6}
&\sum_{\lambda^{\prime \prime }\lambda^{\prime \prime \prime }}Z^{1/2}_{\lambda\lambda^{\prime \prime }}(\mathbf{k})
G^{-1}_{\lambda^{\prime \prime }\lambda^{\prime \prime \prime }}(\mathbf{k}%
;\omega)Z^{1/2}_{\lambda^{\prime \prime \prime }\lambda^{\prime }}(\mathbf{k}%
)= i\omega\delta_{\lambda\lambda^{\prime }}+  \notag \\
&\sum_{\lambda^{\prime \prime }\lambda^{\prime \prime \prime
}}Z^{1/2}_{\lambda\lambda^{\prime \prime }}(\mathbf{k}) [(\mu-\varepsilon^{%
\mathbf{k}}_{\lambda^{\prime \prime }})\delta_{\lambda^{\prime \prime }\lambda^{\prime \prime \prime }} -\Sigma^{c}_{\lambda^{\prime \prime
}\lambda^{\prime \prime \prime }}(\mathbf{k};0)]Z^{1/2}_{\lambda^{\prime \prime \prime }\lambda^{\prime }}(\mathbf{k}).
\end{align}

The second term on the right hand side of this equation is a Hermitian matrix, the quasiparticle  Hamiltonian matrix.  It is  diagonalized in subroutine \verb"BANDS_QP".

\begin{align}  \label{qp_7}
&\mu\delta_{\lambda\lambda^{\prime }}-H^{\mathbf{k}}_{\lambda\lambda^{\prime
}}  \notag \\
&=\sum_{\lambda^{\prime \prime }\lambda^{\prime \prime \prime
}}Z^{1/2}_{\lambda\lambda^{\prime \prime }}(\mathbf{k}) [(\mu-\varepsilon^{%
\mathbf{k}}_{\lambda^{\prime \prime }})\delta_{\lambda^{\prime \prime }\lambda^{\prime \prime \prime }} -\Sigma^{c}_{\lambda^{\prime \prime
}\lambda^{\prime \prime \prime }}(\mathbf{k};0)]Z^{1/2}_{\lambda^{\prime
\prime \prime }\lambda^{\prime }}(\mathbf{k})  \notag \\
&=\sum_{i}Q^{\mathbf{k}}_{\lambda i}E^{\mathbf{k}}_{i}Q^{^{\dagger}\mathbf{k}%
}_{i\lambda^{\prime }},
\end{align}
where $E_{i}^{\mathbf{k}}$ are the effective eigenvalues. After that, we can rewrite (\ref{qp_6}) as follows

\begin{align}  \label{qp_8}
\sum_{\lambda^{\prime \prime }\lambda^{\prime \prime \prime }}Z^{1/2}_{\lambda\lambda^{\prime \prime }}(\mathbf{k})
&G^{-1}_{\lambda^{\prime \prime }\lambda^{\prime \prime \prime }}(\mathbf{k}%
;\omega)Z^{1/2}_{\lambda^{\prime \prime \prime }\lambda^{\prime }}(\mathbf{k}%
)  \notag \\
&= \sum_{i}Q^{\mathbf{k}}_{\lambda i}\big[i\omega+\mu-E^{\mathbf{k}}_{i}\big]%
Q^{^{\dagger}\mathbf{k}}_{i\lambda^{\prime }},
\end{align}
or, for the Green function

\begin{align}  \label{qp_9}
G^{\mathbf{k}}_{\lambda\lambda^{\prime }}(\omega)= \sum_{i}\frac{(Z^{1/2}Q)^{%
\mathbf{k}}_{\lambda i}(Q^{\dagger}Z^{1/2})^{\mathbf{k}}_{i\lambda^{\prime }}%
}{i\omega+\mu-E^{\mathbf{k}}_{i}}.
\end{align}

This expression differs from the full GW Greens function  by a linear approximation for the
frequency dependent self energy.

At this point, we  construct the quasiparticle Greens funcion 
(step (\ref{GW_5}) in section \ref{GW})  by  
setting  $Z_{\lambda \lambda ^{\prime }}^{\mathbf{k}%
}=\delta _{\lambda \lambda ^{\prime }}$ in the above equation to obtain

\begin{align}  \label{qp_10}
G^{\mathbf{k}}_{\lambda\lambda^{\prime }}(\omega)= \sum_{i}\frac{Q^{\mathbf{k%
}}_{\lambda i}Q^{^{\dagger}\mathbf{k}}_{i\lambda^{\prime }}}{i\omega+\mu-E^{%
\mathbf{k}}_{i}}.
\end{align}

\section{Polarizability calculation and scaling}
\label{P_R}

In accordance with the MT-geometry there are three essentially different contributions to the polarizability corresponding to i)when both space arguments of $P$ belong to
MT spheres (Mt-Mt); ii) one of them belongs to a MT-sphere and another belongs to the Interstitial region (Mt-Int); iii) both arguments belong to the
interstitial region (Int-Int). Below we consider three cases separately, describing how the Green function is transformed from the band representation
to the real space, how we calculate the polarizability, and how we transform it from the real space to the reciprocal space.

Our parallelization strategy here is to use two-dimensional grid of MPI-processes. The first MPI-dimension in polarizability calculations is
associated with $\tau$-variable with each process doing calculation only on its own set of $\tau$-indexes. It is most efficient because all formulae in this section are totally independent for different $\tau$'s. The second
dimension of MPI grid is used whenever it is appropriate as described briefly below. Namely, every process associated with the second dimension of the MPI grid is carrying out calculations on its own set of $\mathbf{k}$ points, or on its own set of triplets ($\mathbf{Rtt}'$).

In the following sections we will present the scalings associated with the principle steps of the algorithm. For convenience we summarize main notations here: $N_{at}$ is the number of atoms in the unit cell; $N_{orb}$ is the number of orbitals per atom in the LAPW+LO representation (for typical numbers see Table \ref{Time} below). The number of bands is approximately equal to $N_{at}N_{orb}$, so we will not use the number of bands below. Further, the number of plane waves in the interstitial region used to represent the fermionic functions approximately equals the number of bands, so we do not use it below as well. Bosonic functions make the major impact on the calculation time. So, it is practical to take into account their numbers more carefully. $N^{Mt}_{pb}$ is the number of product basis orbitals inside MT-sphere (per atom); $N^{Int}_{pb}$ is the number of plane waves associated with product basis in the interstitial region (per atom); $N_{r}$ is the number of points in the regular real space mesh in the unit cell (per atom); $N_{k}$ is the number of points in the whole Brillouin zone; $N_{\tau}$ is the number of points in $\tau$-mesh. The number of points in the fermionic and bosonic frequency meshes is about the same as the number of $\tau$-points, so we use the latter in all cases.

\subsection{Mt-Mt part of polarizability} \label{P_MM}

When both space arguments belong to MT-spheres, real space representation means that we represent $G$ as an expansion in local orbitals inside the
spheres at $\mathbf{t}$ and $\mathbf{t}'$ in the unit cells separated by translation vector $\mathbf{R}$

\begin{align}  \label{GR_1}
G^{\alpha}(\mathbf{r};\mathbf{r}';\tau)|_{\mathbf{t}+\mathbf{R};\mathbf{t}'}=\sum_{LL'}\varphi^{\alpha\mathbf{t}}_{L}(\mathbf{r})
G_{\mathbf{t}L;\mathbf{t}'L'}^{\alpha\mathbf{R}}(\tau )\varphi^{\alpha\mathbf{t}'}_{L'}(\mathbf{r}'),
\end{align}
with the coefficients found with (\ref{LAPW_2}):

\begin{align}  \label{GR_2}
G_{\mathbf{t}L;\mathbf{t}'L'}^{\alpha\mathbf{R}}(\tau )=\frac{1}{N_{ \mathbf{k}}}\sum_{\mathbf{k}}e^{i\mathbf{k}\mathbf{R}} \sum_{\lambda
\lambda'}Z_{\mathbf{t}L}^{\alpha\mathbf{k}\lambda}G_{\lambda \lambda'}^{\alpha\mathbf{k}}(\tau)Z_{\mathbf{t}'L'}^{^{*}\alpha\mathbf{k}\lambda'}.
\end{align}

So the first step in Mt-Mt case is to transform Green's function from band representation to the representation (\ref{GR_1}) using (\ref{GR_2}). The scaling associated with the evaluation of (\ref{GR_2}) is $\left[(N_{at}N_{orb})^{3}N_{k}+(N_{at}N_{orb})^{2}N_{k}\ln N_{k}\right]N_{\tau}$. The first term corresponds to the sum over ($\lambda,\lambda'$) indexes which scales as $(N_{at}N_{orb})^{3}$ for every $\mathbf{k}$-point and $\tau$. The second term is related to the fast Fourier transform from $\mathbf{k}$ space to the $\mathbf{R}$ space which scales as $N_{k}\ln N_{k}$ for each matrix element and $\tau$.
We use second dimension of MPI-grid to calculate matrix products (sums over band indexes) spreading different $\mathbf{k}$'s over the MPI
processes. Then we switch MPI parallelization to perform FFT for different indexes ($\mathbf{t}'L'$). In the code the above Green function
transformation is performed in \verb"G_RS_FROM_KS_MM" subroutine.

The expression for the polarizability then follows from (\ref{GR_1}) and (\ref{GW_1})

\begin{align}  \label{P_MM_1}
P_{\mathbf{t}K;\mathbf{t}'K'}^{\mathbf{R}}(\tau)=&-\sum_{\alpha} \sum_{LL''}\langle M_{K}^{\mathbf{t}}|\varphi_{L}^{\alpha\mathbf{t}}\varphi
_{L''}^{\alpha\mathbf{t}}\rangle \nonumber
\\
& \times \sum_{L'}G_{\mathbf{t}L;\mathbf{t}'L'}^{\alpha\mathbf{R}}(\tau )\sum_{L'''}
G_{\mathbf{t}L'';\mathbf{t}'L'''}^{\alpha;\mathbf{R}}(\beta -\tau)  \nonumber \\
&\times \langle \varphi_{L'}^{\alpha\mathbf{t}'}\varphi_{L'''}^{\alpha\mathbf{t}'}|M_{K'}^{\mathbf{t}'}\rangle ,
\end{align}

The scaling of (\ref{P_MM_1}) is $\left[2N_{orb}^{3}N^{Mt}_{pb}+(N_{orb}N_{pb}^{Mt})^{2}\right]N_{at}^{2}N_{k}N_{\tau}$. To evaluate (\ref{P_MM_1}) we use second MPI dimension to parallelize the triplets ($\mathbf{R};\mathbf{t};\mathbf{t}'$). In the code
(\ref{P_MM_1}) is implemented in \verb"P_MM_R" subroutine.

Transform to the reciprocal space consists in one FFT transform.

\begin{align}  \label{P_TR_1}
P_{\mathbf{t}K;\mathbf{t}'K'}^{\mathbf{q}}(\tau)=\sum_{\mathbf{R}}e^{-i\mathbf{q}\mathbf{R}}P_{\mathbf{t}K;\mathbf{t}'K'}^{\mathbf{R}}(\tau),
\end{align}
which is implemented in the subroutine \verb"P_MM_Q_FROM_R". The scaling of (\ref{P_TR_1}) is $(N_{pb}^{Mt}N_{at})^{2}N_{k}\ln N_{k}N_{\tau}$.

\subsection{Mt-Int part of polarizability} \label{P_MI}

In this case the second space argument in Eq.(\ref{GR_3}) runs over the regular $\mathbf{r}$-mesh in the whole unit cell, whereas for the first space argument we use an expansion in local
orbitals:

\begin{align}  \label{GR_3}
G^{\alpha}(\mathbf{r};\mathbf{r}';\tau)|_{\mathbf{r}\in\mathbf{t}+\mathbf{R}}=\sum_{L}\varphi^{\alpha\mathbf{t}}_{L}(\mathbf{r})
G_{\mathbf{t}L;\mathbf{r}'}^{\alpha\mathbf{R}}(\tau ).
\end{align}

The corresponding coefficients $G_{\mathbf{t}L;\mathbf{r}'}^{\alpha\mathbf{R}}(\tau)$ are obtained in two steps (subroutine
\verb"G_RS_FROM_KS_MI"):

\begin{align}  \label{GR_5}
G_{\mathbf{t}L;\mathbf{G}'}^{\alpha\mathbf{k}}(\tau)=\frac{1}{\sqrt{\Omega_{0}}}\sum_{\lambda \lambda'}
Z_{\mathbf{t}L}^{\alpha\mathbf{k}\lambda}G_{\lambda \lambda'}^{\alpha\mathbf{k}}(\tau) A_{\mathbf{G}'}^{^{*}\alpha\mathbf{k}\lambda'},
\end{align}
and
\begin{align}  \label{GR_6}
G^{\alpha\mathbf{R}}_{\mathbf{t}L;\mathbf{r}'}(\tau )=\frac{1}{N_{\mathbf{k}}}\sum_{\mathbf{k}}e^{i\mathbf{k}\mathbf{R}}
\sum_{\mathbf{G}'}e^{-i(\mathbf{k}+\mathbf{G}')\mathbf{r}'} G_{\mathbf{t}L;\mathbf{G}'}^{\alpha\mathbf{k}}(\tau ).
\end{align}

The scaling of (\ref{GR_5}) is $N_{at}^{3}N_{orb}^{3}N_{k}N_{\tau}$. The scaling associated with the evaluation of (\ref{GR_6}) is $N_{at}^{2}N_{orb}N_{r}N_{k}N_{\tau}\left[\ln(N_{at}N_{r})+\ln N_{k}\right]$. MPI-parallelization is used in (\ref{GR_5}) and (\ref{GR_6}) to perform calculations for different $\mathbf{k}$'s independently.

The expression for the polarizability follows from (\ref{GR_3}) and (\ref{GW_1})

\begin{align}  \label{PR_2}
P^{\mathbf{R}}_{\mathbf{t}K;\mathbf{r}'}(\tau)=-\sum_{\alpha}\sum_{LL'} \langle
M^{\mathbf{t}}_{K}|\varphi^{\alpha\mathbf{t}}_{L}\varphi^{\mathbf{t}}_{L'}\rangle
G^{\alpha\mathbf{R}}_{\mathbf{t}L;\mathbf{r}'}(\tau)G^{\alpha\mathbf{R}}_{\mathbf{t}L';\mathbf{r}'}(\beta-\tau).
\end{align}

The scaling of (\ref{PR_2}) is $(N_{at}N_{orb})^{2}N^{Mt}_{pb}N_{r}N_{k}N_{\tau}$. MPI-parallelization is used in (\ref{PR_2}) to perform the calculations for different $\mathbf{R}$'s independently (subroutine \verb"P_IM_R").

The reciprocal space representation in original product basis is obtained with two FFTs:

\begin{align}  \label{P_TR_2}
\widetilde{P}^{\mathbf{q}}_{\mathbf{t}K;\mathbf{G}'}(\tau)=\frac{1}{N_{\mathbf{r}}}\sum_{\mathbf{r}'}e^{i(\mathbf{q}+\mathbf{G}')\mathbf{r}'}
\sum_{\mathbf{R}}e^{-i\mathbf{q}\mathbf{R}}P_{\mathbf{t}K;\mathbf{r}'}^{\mathbf{R}}(\tau).
\end{align}

The scaling of (\ref{P_TR_2}) is $N_{at}^{2}N_{k}N_{\tau}N^{Mt}_{pb}N_{r}\left[\ln N_{k}+\ln (N_{at}N_{r})\right]$. Representation in the dual basis is obtained after an additional step

\begin{align}  \label{P_TR_2A}
P^{\mathbf{q}}_{\mathbf{t}K;K'}(\tau)=\sum_{\mathbf{G}'}\widetilde{P}^{\mathbf{q}}_{\mathbf{t}K;\mathbf{G}'}(\tau)\langle
e^{i(\mathbf{q}+\mathbf{G}')\mathbf{r}'}|M^{\mathbf{q}}_{K'}\rangle_{Int},
\end{align}
where $\langle
e^{i(\mathbf{q}+\mathbf{G}')\mathbf{r}'}|M^{\mathbf{q}}_{K'}\rangle_{Int}$ represents the integral of the product of two plane waves over the interstitial region and it is done analytically.

The scaling of (\ref{P_TR_2A}) is $N_{at}^{3}N_{k}N_{\tau}N^{Mt}_{pb}N_{pb}^{^{2}Int}$. MPI-parallelization is used in (\ref{P_TR_2}) and (\ref{P_TR_2A}) to perform calculations for different $\mathbf{q}$'s independently (subroutine
\verb"P_IM_Q_FROM_R").

\subsection{Int-Int part of polarizability} \label{P_II}

In this case both space arguments run over the regular mesh in the unit cell. The real space representation for $G$ is obtained in two steps (subroutines
\verb"G_K_G_R1" and \verb"G_RR_R_R1_STAR"):

\begin{align}  \label{GR_3A}
G_{\mathbf{G};\mathbf{G}'}^{\alpha\mathbf{k}}(\tau )=\frac{1}{\Omega_{0}}\sum_{\lambda \lambda'}
A_{\mathbf{G}}^{\alpha\mathbf{k}\lambda}G_{\lambda \lambda'}^{\alpha\mathbf{k}}(\tau)A_{\mathbf{G}'}^{^{*}\alpha\mathbf{k}\lambda'},
\end{align}
and
\begin{align}  \label{GR_4}
G^{\alpha\mathbf{R}}_{\mathbf{r};\mathbf{r}'}(\tau )=
\frac{1}{N_{\mathbf{k}}}\sum_{\mathbf{k}}e^{i\mathbf{k}\mathbf{R}}\sum_{\mathbf{G};\mathbf{G}'}e^{i(\mathbf{k}+\mathbf{G})\mathbf{r}}
G_{\mathbf{G};\mathbf{G}'}^{\alpha\mathbf{k}}(\tau )e^{-i(\mathbf{k}+\mathbf{G}')\mathbf{r}'}.
\end{align}

The scalings of (\ref{GR_3A}) and (\ref{GR_4}) are $(N_{at}N_{orb})^{3}N_{k}N_{\tau}$ and $(N_{at}N_{r})^{2}N_{k}N_{\tau}\left[2\ln (N_{at}N_{r})+\ln N_{k}\right]$ correspondingly. MPI-parallelization is used in (\ref{GR_3A}) and (\ref{GR_4}) to perform calculations for different $\mathbf{k}$'s independently.

The formula for the polarizability is very simple in this case

\begin{align}  \label{PR_3}
P^{\mathbf{R}}_{\mathbf{r}\mathbf{r}'}(\tau)=-\sum_{\alpha}G^{\alpha\mathbf{R}}_{\mathbf{r}\mathbf{r}'}(\tau)
G^{\alpha\mathbf{R}}_{\mathbf{r}\mathbf{r}'}(\beta-\tau).
\end{align}

The scaling of (\ref{PR_3}) is $(N_{at}N_{r})^{2}N_{k}N_{\tau}$. In (\ref{PR_3}) we use MPI processes associated with index $\mathbf{r}'$ and $\tau$.

The reciprocal space representation in the original product basis is obtained with three FFTs:

\begin{align}  \label{P_TR_3}
\widetilde{P}^{\mathbf{q}}_{\mathbf{G}\mathbf{G}'}(\tau)&=\frac{1}{N_{\mathbf{r}}}\sum_{\mathbf{r}}e^{i(\mathbf{q}+\mathbf{G})\mathbf{r}}
\frac{1}{N_{\mathbf{r}}}\sum_{\mathbf{r}'}e^{-i(\mathbf{q}+\mathbf{G}')\mathbf{r}'}  \nonumber \\
&\times\sum_{\mathbf{R}}e^{-i\mathbf{q}\mathbf{R}}P^{\mathbf{R}}_{\mathbf{r};\mathbf{r}'}(\tau).
\end{align}

The scalings of Eq. (\ref{P_TR_3}) is $N_{at}^{2}N_{r}N_{k}N_{\tau}\left[(N_{r}+N_{pb}^{Int})\ln (N_{at}N_{r})+N_{r}\ln N_{k}\right]$. Representation in the dual basis follows as additional matrix multiplications

\begin{align}  \label{P_TR_3A}
P^{\mathbf{q}}_{K;K'}(\tau)=\sum_{\mathbf{G}\mathbf{G}'}\langle
e^{i(\mathbf{q}+\mathbf{G})\mathbf{r}}|M^{\mathbf{q}}_{K}\rangle^{*}_{Int}\widetilde{P}^{\mathbf{q}}_{\mathbf{G}\mathbf{G}'}(\tau)\langle
e^{i(\mathbf{q}+\mathbf{G}')\mathbf{r}'}|M^{\mathbf{q}}_{K'}\rangle_{Int}.
\end{align}

The scaling of (\ref{P_TR_3A}) is $(N_{at}N_{pb}^{Int})^{3}N_{k}N_{\tau}$. MPI-parallelization is used in (\ref{P_TR_3}) and (\ref{P_TR_3A}) to perform calculations for different $\mathbf{q}$'s independently. Formulae
(\ref{PR_3})-(\ref{P_TR_3A}) are implemented in the subroutine \verb"P_II_SOLID".

\section{Screened interaction}
\label{W_bas}

Equation (\ref{GW_2}) in reciprocal space reads as the following

\begin{align}  \label{W_1}
W^{\mathbf{q}}_{KK'}(\nu)=V^{\mathbf{q}}_{KK'}+\sum_{K''K'''} V^{\mathbf{q}}_{KK''}P^{\mathbf{q}}_{K''K'''}(\nu)W^{\mathbf{q}}_{K'''K'}(\nu).
\end{align}

The scaling of (\ref{W_1}) is $(N_{at}[N_{pb}^{Mt}+N_{pb}^{Int}])^{3}N_{k}N_{\tau}$. We associate two-dimensional mesh of MPI-processes with variables $\mathbf{q}$ and $\nu$. Formula (\ref{W_1}) is implemented in the subroutine
\verb"WS_K_NU_SOLID_0".

\section{Dynamic self energy}
\label{sigma}

According to the division of the screened interaction into bare Coulomb V and dynamic part $\widetilde{W}$ ($W=V+\widetilde{W}$) the self energy is also divided into static and dynamic. Here we consider the evaluation of the dynamic part only. Static part is evaluated similarly with obvious simplifications in the formulae.

In accordance with MT-geometry there are three essentially different contributions to the self energy corresponding to i)when both space arguments of belong to
MT spheres (Mt-Mt); ii) one of them belongs to a MT-sphere and another belongs to the Interstitial (Mt-Int); iii) both arguments belong to the
interstitial region (Int-Int). Below we consider three cases separately, describing how the screened interaction is transformed from reciprocal
space to the real space, how we calculate the dynamic self energy, and how we transform it from real space back to reciprocal space and band
representation.

Our parallelization strategy here is similar to the strategy in polarizability calculations.

\subsection{Mt-Mt part of self energy} \label{S_MM}

When both space arguments belong to MT-spheres, real space representation means that we represent $\widetilde{W}$ as an expansion in product basis
functions inside the spheres at $\mathbf{t}$ and $\mathbf{t}'$ in the unit cells separated by translation vector $\mathbf{R}$

\begin{align}  \label{S_MM_1}
\widetilde{W}_{\mathbf{t}K;\mathbf{t}'K'}^{\mathbf{R}}(\tau)=\frac{1}{N_{\mathbf{k}}}\sum_{\mathbf{q}}e^{i\mathbf{q}\mathbf{R}}
\widetilde{W}_{\mathbf{t}K;\mathbf{t}'K'}^{\mathbf{q}}(\tau).
\end{align}

We use the second dimension of MPI-grid to calculate matrix products (sums over band indexes) spreading different $\mathbf{q}$'s over the MPI
processes. Formula (\ref{S_MM_1}) is implemented in the subroutine \verb"W_MM". The scaling of (\ref{S_MM_1}) is $(N_{at}N_{pb}^{Mt})^{2}N_{\tau}N_{k}\ln N_{k}$.

The expression for the self energy follows from (\ref{GR_1}) and (\ref{GW_3}) (subroutine \verb"SIGC_MM_R")

\begin{align}  \label{S_MM_2}
\Sigma_{\mathbf{t}L;\mathbf{t}'L'}^{\alpha\mathbf{R}}(\tau)=&-\sum_{L''L'''}\sum_{KK'}\langle \varphi_{L}^{\alpha\mathbf{t}}|\varphi
_{L''}^{\alpha\mathbf{t}}M_{K}^{\mathbf{t}}\rangle \nonumber
\\
& \times G_{\mathbf{t}L;\mathbf{t}'L'}^{\alpha\mathbf{R}}(\tau )
\widetilde{W}_{\mathbf{t}K;\mathbf{t}'K'}^{\mathbf{R}}(\beta-\tau)  \nonumber \\
&\times \langle \varphi_{L'}^{\alpha\mathbf{t}'}|\varphi_{L'''}^{\alpha\mathbf{t}'}M_{K'}^{\mathbf{t}'}\rangle ,
\end{align}

The scaling of (\ref{S_MM_2}) is $\left[2N_{orb}^{3}N^{Mt}_{pb}+(N_{orb}N_{pb}^{Mt})^{2}\right]N_{at}^{2}N_{k}N_{\tau}$. To evaluate (\ref{S_MM_2}) we use second MPI dimension to parallelize the triplets ($\mathbf{R};\mathbf{t};\mathbf{t}'$).

Transform to the reciprocal space consists in one FFT transform (subroutine \verb"SIGC_MM_K_FROM_R").

\begin{align}  \label{S_TR_1}
\Sigma_{\mathbf{t}L;\mathbf{t}'L'}^{\mathbf{k}}(\tau)=\sum_{\mathbf{R}}e^{-i\mathbf{k}\mathbf{R}}\Sigma_{\mathbf{t}L;\mathbf{t}'L'}^{\mathbf{R}}(\tau),
\end{align}
which scales as $(N_{at}N_{orb})^{2}N_{\tau}N_{k}\ln N_{k}$.

\subsection{Mt-Int part of self energy} \label{S_MI}

In this case the transform of $W$ to real space involves two FFT's (subroutine \verb"V_IM_R_FROM_K"):

\begin{align}  \label{W_RT_2}
\widetilde{W}^{\mathbf{R}}_{\mathbf{t}K;\mathbf{r}}(\tau)= \frac{1}{N_{\mathbf{k}}}\sum_{\mathbf{q}}e^{i\mathbf{q}\mathbf{R}}
\sum_{\mathbf{G}}e^{-i(\mathbf{q}+\mathbf{G})\mathbf{r}}\widetilde{W}_{\mathbf{t}K;\mathbf{G}}^{\mathbf{q}}(\tau).
\end{align}

The scaling of (\ref{W_RT_2}) is $N_{at}^{2}\ln (N_{at}N_{r}) N_{pb}^{Mt}N_{r}N_{k}N_{\tau}$. MPI-parallelization is used in (\ref{W_RT_2}) to perform calculations for different $\mathbf{q}$'s independently.

The expression for the self energy follows from (\ref{GR_3}) and (\ref{GW_3})

\begin{align}  \label{S_MI_2}
\Sigma^{\alpha\mathbf{R}}_{\mathbf{t}L;\mathbf{r}'}(\tau)=-\sum_{L'K} \langle
\varphi^{\alpha\mathbf{t}}_{L}|\varphi^{\mathbf{t}}_{L'}M^{\mathbf{t}}_{K}\rangle
G^{\alpha\mathbf{R}}_{\mathbf{t}L;\mathbf{r}'}(\tau)\widetilde{W}^{\mathbf{R}}_{\mathbf{t}K;\mathbf{r}'}(\beta-\tau),
\end{align}
which scales as $N_{at}^{2}N_{orb}^{2}N_{pb}^{Mt}N_{r}N_{k}N_{\tau}$.

MPI-parallelization is used in (\ref{S_MI_2}) to perform calculations for different $\mathbf{R}$'s independently. Formula (\ref{S_MI_2}) is
implemented in the subroutine \verb"SIGC_IM_R".

Transformation to the band states representation is achieved in a few steps. They are implemented in the subroutine \verb"SIGC_IM_K_FROM_R".

First we apply FFT

\begin{equation}  \label{S_MI_3}
\Sigma^{\alpha\mathbf{k}}_{\mathbf{t}L;\mathbf{r}'}(\tau)=\sum_{\mathbf{R}}e^{-i\mathbf{k}\mathbf{R}}
\Sigma^{\alpha\mathbf{R}}_{\mathbf{t}L;\mathbf{r}'}(\tau),
\end{equation}
with scaling $N_{at}^{2}N_{orb}N_{r}N_{\tau}N_{k}\ln N_{k}$.

At this point the function is represented by its values at the homogeneous $\mathbf{r}'$-mesh in the whole unit cell. In order to perform
integration over the interstitial region we again apply FFT to transform it into equivalent linear combination of plane waves

\begin{align}  \label{S_MI_4}
\Sigma^{\alpha\mathbf{k}}_{\mathbf{t}L;\mathbf{r}'}(\tau)=\sum_{\mathbf{G}'} \tilde{\Sigma}^{\alpha\mathbf{k}}_{\mathbf{t}L;\mathbf{G}'}
e^{-i(\mathbf{k}+\mathbf{G}')\mathbf{r}'},
\end{align}
with the coefficients
\begin{align}  \label{S_MI_5}
\tilde{\Sigma}^{\alpha\mathbf{k}}_{\mathbf{t}L;\mathbf{G}'}(\tau)=\frac{1}{N_{\mathbf{r}}}\sum_{\mathbf{r}'}
e^{i(\mathbf{k}+\mathbf{G}')\mathbf{r}'}\Sigma^{\alpha\mathbf{k}}_{\mathbf{t}L;\mathbf{r}'}(\tau).
\end{align}
The scaling of (\ref{S_MI_5}) is $N_{at}^{2}N_{orb}N_{r}N_{k}N_{\tau}\ln (N_{at}N_{r})$. The form (\ref{S_MI_4}) allows us to integrate over the interstitial region analytically and we obtain
\begin{align}  \label{S_MI_6}
\Sigma^{\alpha\mathbf{k}}_{\mathbf{t}L;\mathbf{G}'}(\tau)=\frac{1}{\sqrt{\Omega_{0}}}
\sum_{\mathbf{G}''}\tilde{\Sigma}^{\alpha\mathbf{k}}_{\mathbf{t}L;\mathbf{G}''}(\tau)S^{\mathbf{k}}_{\mathbf{G}''\mathbf{G}'},
\end{align}
which scales as $N_{at}^{3}N_{orb}^{2}N_{pb}^{Int}N_{k}N_{\tau}$. $S^{\mathbf{k}}_{\mathbf{G}''\mathbf{G}'}$ in (\ref{S_MI_6}) is the integral of the product of two plane waves ($e^{-i(\mathbf{k}+\mathbf{G}'')\mathbf{r}}$ and $e^{i(\mathbf{k}+\mathbf{G}''')\mathbf{r}}$)  taken over the interstitial region.

Finally, the contribution to the band state representation follows

\begin{equation}  \label{S_MI_7}
\Sigma^{\alpha\mathbf{k}}_{\lambda\lambda'}(\tau)|^{Mt}_{Int}=\sum_{\mathbf{t}L}\sum_{\mathbf{G}'} Z^{^{*}\alpha\mathbf{k}\lambda}_{\mathbf{t}L}
\Sigma^{\alpha\mathbf{k}}_{\mathbf{t}L;\mathbf{G}'}(\tau)A^{\alpha\mathbf{k}\lambda'}_{\mathbf{G}'}+H.C..
\end{equation}

The scaling of (\ref{S_MI_7}) is $(N_{a}N_{orb})^{3}N_{k}N_{\tau}$. MPI-parallelization is used in (\ref{S_MI_3}-\ref{S_MI_7}) to perform calculations for different $\mathbf{k}$'s independently.

\subsection{Int-Int part of self energy} \label{S_II}

In this case both space arguments run over the regular mesh in the unit cell. Real space representation for $W$ is obtained with three FFT's
(subroutines \verb"W_Q_G_R1" and \verb"W_RR_R_R1_STAR"):

\begin{align}  \label{S_II_1}
\widetilde{W}^{\mathbf{R}}_{\mathbf{r};\mathbf{r}'}(\tau)&=\frac{1}{N_{\mathbf{k}}}\sum_{\mathbf{q}}e^{i\mathbf{q}\mathbf{R}} \sum_{\mathbf{G}}
e^{-i(\mathbf{q}+\mathbf{G})\mathbf{r}} \sum_{\mathbf{G}'}e^{i(\mathbf{q}+\mathbf{G}')\mathbf{r}'}  \nonumber \\
&\times \widetilde{W}^{\mathbf{q}}_{\mathbf{G}\mathbf{G}'}(\tau),
\end{align}
with scaling    $N_{at}^{2}N_{r}N_{k}N_{\tau}\nonumber\\ \times\left[N_{pb}^{Int}\ln (N_{at}N_{r})+N_{r}\ln (N_{at}N_{r}) + N_{r}\ln N_{k}\right]$.

MPI-parallelization is used in (\ref{S_II_1}) to perform calculations for different $\mathbf{k}$'s independently.

The formula for the self energy is very simple in this case

\begin{equation}  \label{S_II_2}
\Sigma^{\alpha\mathbf{R}}_{\mathbf{r}\mathbf{r}'}=-G^{\alpha\mathbf{R}}_{\mathbf{r}\mathbf{r}'}(\tau )
\widetilde{W}^{\mathbf{R}}_{\mathbf{r}\mathbf{r}'}(\beta -\tau ).
\end{equation}

The scaling of (\ref{S_II_2}) is $(N_{at}N_{r})^{2}N_{\tau}N_{k}$. In (\ref{S_II_2}) we use MPI processes associated with index $\mathbf{r}'$.

Then we apply FFT
\begin{equation}  \label{S_II_3}
\Sigma^{\alpha\mathbf{k}}_{\mathbf{r}\mathbf{r}'}(\tau)=\sum_{\mathbf{R}}e^{-i\mathbf{k}\mathbf{R}}
\Sigma^{\alpha\mathbf{R}}_{\mathbf{r}\mathbf{r}'}(\tau).
\end{equation}
The scaling of (\ref{S_II_3}) is $(N_{at}N_{r})^{2}N_{\tau}N_{k}\ln N_{k}$. Similar to the MT-Int case, we use FFT to transform it into equivalent linear combination of plane waves

\begin{align}  \label{S_II_4}
\Sigma^{\alpha\mathbf{k}}_{\mathbf{r}\mathbf{r}'}(\tau)=\sum_{\mathbf{G}}\sum_{\mathbf{G}'}
e^{i(\mathbf{k}+\mathbf{G})\mathbf{r}}\tilde{\Sigma}^{\alpha\mathbf{k}}_{\mathbf{G};\mathbf{G}'}(\tau)e^{-i(\mathbf{k}+\mathbf{G}')\mathbf{r}'},
\end{align}
with the coefficients
\begin{align}  \label{S_II_5}
\tilde{\Sigma}^{\alpha\mathbf{k}}_{\mathbf{G};\mathbf{G}'}(\tau)=\frac{1}{N^{2}_{\mathbf{r}}}\sum_{\mathbf{r}\mathbf{r}'}
e^{-i(\mathbf{k}+\mathbf{G})\mathbf{r}}e^{i(\mathbf{k}+\mathbf{G}')\mathbf{r}'}\Sigma^{\alpha\mathbf{k}}_{\mathbf{r}\mathbf{r}'}(\tau).
\end{align}
The scaling of (\ref{S_II_5}) is $N_{at}^{2}N_{r}\ln(N_{at}N_{r})N_{k}N_{\tau}\left[N_{r}+N_{pb}^{Int}\right]$. The form (\ref{S_II_4}) allows us to integrate over the interstitial region analytically and as a result we obtain
\begin{align}  \label{S_II_6}
\Sigma^{\alpha\mathbf{k}}_{\mathbf{G};\mathbf{G}'}(\tau)=\frac{1}{\Omega_{0}}
\sum_{\mathbf{G}''\mathbf{G}'''}S^{\mathbf{k}}_{\mathbf{G}\mathbf{G}''} \tilde{\Sigma}^{\alpha\mathbf{k}}_{\mathbf{G}'';\mathbf{G}'''}(\tau)
S^{\mathbf{k}}_{\mathbf{G}'''\mathbf{G}'},
\end{align}
with scaling $N_{at}^{3}N_{pb}^{Int}N_{orb}N_{k}N_{\tau}\left[N_{pb}^{Int}+N_{orb}\right]$.

Finally, the contribution to the band state representation from the interstitial is given by

\begin{equation}  \label{S_II_7}
\Sigma^{\alpha\mathbf{k}}_{\lambda\lambda'}(\tau)|^{Int}_{Int}=\sum_{\mathbf{G}\mathbf{G}'} A^{^{*}\alpha\mathbf{k}\lambda}_{\mathbf{G}}
\Sigma^{\alpha\mathbf{k}}_{\mathbf{G};\mathbf{G}'}(\tau)A^{\alpha\mathbf{k}\lambda'}_{\mathbf{G}'}.
\end{equation}

The scaling of (\ref{S_II_7}) is $(N_{at}N_{orb})^{3}N_{k}N_{\tau}$. MPI-parallelization is used in (\ref{S_II_3}-\ref{S_II_7}) to perform calculations for different $\mathbf{k}$'s independently. Formulae
(\ref{S_II_2})-(\ref{S_II_7}) have been implemented in the subroutine \verb"SIGC_II_SOLID".

\section{Results}
\label{res}

In this section we show  how our linearized version of scQPGW performs and compare  the results  to other (non-linearized) implementations of the scQPGW method  in other basis sets  and to   experimental data.

In the Table\ref{B_Gaps} we present our calculated band gaps for selected semiconductors and insulators obtained with linearized scQPGW and compare them with previous
scQPGW calculations and experiment. As one can see,   our results are pretty close to the non-linearized scQPGW results
and systematically overestimate the band gaps. The overestimation generally is in 10-24\% range for all studied materials, excluding antiferromagnetically ordered NiO (error is only 3.9\%) and the f-eletron compound CeO$_{2}$ where the error is large (70\%). Let us also mention that other QPscGW methods produce large error as well for the current material.

Table \ref{B_Widths} presents the band widths of alkali metals Na and K. Formally, alkali metals belong to the s-materials. But as one can see
from the table the error in calculated band width (20-30\%) is a bit larger  than the error in the calculated band gaps for sp-semiconductors. Also, the
error increases when the density of valence electrons is reduced (when going from Sodium to Potassium). That fact was expected because the electron gas of lower density corresponds to more correlated situation.

\begin{table}[t]
\caption{Band gaps (eV) of selected semiconductors and insulators. Experimental data have been cited from Ref.\cite{prb_92_041115} and
\cite{prl_96_226402}. For the present work results we also include the error (\%) relative to the experiment.} \label{B_Gaps}
\begin{center}
\begin{tabular}{@{}c c c c c c}   & &  &  & Present & \\
   & \cite{prb_76_165106,prl_96_226402} & \cite{prl_99_246403} & \cite{prb_92_041115} & work & Exp.\\
\hline
Si  & 1.23  & 1.41  &1.47  &1.40(14.8\%)  &1.22  \\
SiC  & 2.14  & 2.88  & 2.90 &3.08(22.7\%)  &2.51   \\
C  & 6.52  & 6.18  &6.40  &6.71(14.1\%)  &5.88  \\
GaAs  & 1.93  & 1.85  &1.75  &2.08(23.1\%)  &1.69  \\
ZnO  & 3.87  & 3.8  &4.61  &4.47(24.2\%)  &3.60   \\
NiO  & 4.8  &   & 4.97  &4.47(3.9\%)  &4.3  \\
Cu$_{2}$O  &2.36   &   & 2.65  &2.42(10.0\%)  &2.20   \\
TiO$_{2}$  & 3.78  &   & 4.22  &3.80(22.6\%)  &3.1   \\
SrTiO$_{3}$  & 4.19  &   &  &4.01(21.5\%)  &3.3  \\
CeO$_{2}$  & $\sim$5  &   &  &5.83(70.1\%)  &3-3.5
\end{tabular}
\end{center}
\end{table}

\begin{table}[t]
\caption{Band widths (eV) of alkali metals. Experimental data have been taken from Ref.\cite{prl_60_1558} and Ref.\cite{prb_54_7758}. For the
present work results we also include the error (\%) relative to the experiment.} \label{B_Widths}
\begin{center}
\begin{tabular}{@{}c c c c} &  & Present & \\
   & \cite{prl_96_226402}  & work & Exp.\\
\hline
Na  & 3.0   &3.16(19.2\%)  &2.65  \\
K  &   &2.07(29.4\%)  &1.60
\end{tabular}
\end{center}
\end{table}

\begin{table*}[t]
\caption{Main parameters of the calculations and timings. Time is measured in seconds during one iteration. $N_{PB}$ is the size of product basis. $N_{k}$ is the number of irreducible $\mathbf{k}$-points in the Brillouin zone. $n_{\tau}$ is the number of processes used along the $\tau$-dimension of the MPI grid. $n_{k}$ is the number of processes used along the $k$-dimension of the MPI grid.} \label{Time}
\begin{center}
\begin{tabular}{@{}c c c c c c c c c}   & $N_{PB}$  & $N_{LAPW+LO}$ & $N_{k}$  & $n_{\tau}/n_{k}$  
& P  & W  & $\Sigma$ & G\\
\hline
Si  &910  &129  &$12^{3}$  &24/3  &262  &167  &705   & 4  \\
SiC  &903  &118  &$8^{3}$  &24/3  &89  &140  &345   & 3  \\
C  &973  &140  &$12^{3}$  &24/3  &167  &235  &582   & 3  \\
GaAs  &1116  &158  &$4^{3}$  &24/3  &23  &46  &121   & 0  \\
NiO  &2072  &420  &$4^{3}$  &24/6  &427  &334  &1320   & 57  \\
Cu$_{2}$O  &3390  &794  &$4^{3}$  &24/2  &121  &1067  &451   & 178  \\
TiO$_{2}$  &2552  &496  &$4^{3}$  &24/6  &289  &398  &922   & 26  \\
SrTiO$_{3}$  &2598  &373  &$4^{3}$  &24/5  &85  &223  &304   & 3  \\
CeO$_{2}$  &1361  &225  &$6^{3}$  &24/4  &99  &89  &240   & 2  \\
K  &628  &76  &$8^{3}$  &24/1  &22  &31  &56   & 2  \\
Ni  &475  &68  &$16^{3}$  &16/3  &486  &170  &707   & 10
\end{tabular}
\end{center}
\end{table*}

Table \ref{Time} shows the time which was needed to evaluate the  main quantities (P, W, $\Sigma$, and G) during one iteration. As it can be seen for the materials studied, the calculation of the self energy is the most time consuming. However, increasing the size of the product basis (which is proportional to the number of atoms in the unit cell) will eventually make the evaluation of W the  most time consuming. This is clear from the scaling considerations: whereas many parts of the algorithm scale as $N_{at}^{3}$ the evaluation of W has the biggest prefactor.

\section*{Conclusions}
\label{concl}

We presented an implementation of the scQPGW method in LAPW basis set which scales as $N^{3}$ with the number of atoms. Further improvements of the algorithm for large systems would require a removal of the computational bottleneck which is the matrix inversion in Eq.(\ref{W_1}). 
In its current form, this code can serve as a starting point for further diagrammatic 
many body studies on the Matsubara axis in an all electron basis as
was done for example in  
Ref.\cite{arx_1504_07569}.

\section*{Acknowledgments}
\label{acknow}
This work was   supported by the U.S. Department of energy, Office of Science, Basic
Energy Sciences as apart of the Computational Materials Science Program. We thank Sangkook
Choi for many discussions. 





\bibliographystyle{elsarticle-num}







\end{document}